\documentclass[twocolumn]{aastex6}

\usepackage{natbib}
\begin{document}

\title{Metal enrichment leads to low atmospheric C/O ratios in transiting giant exoplanets}

\author{N\'estor Espinoza\altaffilmark{1,2}, Jonathan J. Fortney\altaffilmark{3}, Yamila Miguel\altaffilmark{4}, 
Daniel Thorngren\altaffilmark{5} \& Ruth Murray-Clay\altaffilmark{3}}

\altaffiltext{1}{Instituto de Astrof\'isica, Facultad de F\'isica, Pontificia Universidad Cat\'olica de Chile, 
Av. Vicu\~na Mackenna 4860, 782-0436 Macul, Santiago, Chile.}
\altaffiltext{2}{Millennium Institute of Astrophysics (MAS), Av. Vicu\~na Mackenna 4860, 782-0436 Macul, Santiago, Chile.}
\altaffiltext{3}{Department of Astronomy and Astrophysics, University of California, Santa Cruz, USA.}
\altaffiltext{4}{Laboratoire Lagrange, UMR 7293, Universit\'e de Nice-Sophia Antipolis, CNRS, Observatorie de la Cote dAzur, Blvd de 
l'Observatorie, CS 34229, 06304 Nice cedex 4, France.}
\altaffiltext{5}{Department of Physics, University of California, Santa Cruz, USA.}

\begin{abstract}

We predict the carbon-to-oxygen (C/O) ratios in the hydrogen-helium envelope and atmospheres of a sample of nearly 50 relatively cool ($T_{\mathrm eq}<$ 1000 K) 
transiting gas giant planets. The method involves planetary envelope metallicity estimates that use the structure models of \citet{thorngren:2015} 
and the disk and planetary accretion model of \citet{oberg:2011}.  We find that nearly all of these planets are strongly metal-enriched which, 
coupled with the fact that solid material is the main deliverer of metals in the protoplanetary disk, implies that the substellar C/O ratios of 
their accreted solid material dominate compared to the enhanced C/O ratio of their accreted gaseous component. We predict that these planets will have atmospheres 
that are typically reduced in their C/O compared to parent star values independent of the assessed formation locations, with C/O $<1$ a nearly 
universal outcome within the framework of the model.  We expect water vapor absorption features to be ubiquitous in the atmospheres of these planets, 
and by extension, other gas giants.

\end{abstract}


\section{Introduction} \label{sec:intro}

Over the past several years there has been a significant push in 
exoplanet characterization to better understand the carbon-to-oxygen 
(C/O) ratios in the atmospheres of gas giant exoplanets.  This has been 
in particular motivated by the connections that can potentially be made 
to the formation location and accretion history of the planets 
\citep[see, e.g.,][]{madhusudhan:2011,oberg:2011,madhusudhan:2012,
mordasini:2016}.  This work is also quite timely given that the 
\emph{Juno} Mission at Jupiter aims to measure the planet's 
atmospheric water abundance via microwave emission, and constrain 
the planet's C/O, which impacts our knowledge of the planet's 
interior structure and formation \citep{helled:2014}.

As is known in our solar system \citep{atreya:2016}, giant 
planet's atmospheres need not take on the composition of their parent stars. 
Instead, the composition of the atmospheres is defined by the accretion of gas and solids, 
with the composition of these species dictated by the position of 
the planet in the protoplanetary disk, which in turn defines how much of each element 
is available in solid and gas form. This idea has been one of the key arguments in 
order to link planet formation and the composition of giant planet atmospheres 
\citep[see, e.g.,][]{madhusudhan:2012,madhusudhan:2011,oberg:2011}.

There may well be additional tracers of formation and accretion history, but C/O ratios have 
driven the attention of the community mainly because of their potentially
large observational effects. For example, water features in the 
infrared weaken or disappear in hot exoplanets for C/O $\gtrsim 1$ according to chemical 
equilibrium calculations, because the water vapor abundance is strongly 
reduced \citep{madhusudhan:2012}. This prediction has motivated several studies that aim 
to detect water vapor, in particular in transmission spectra, because the 
presence of a water feature could then constrain a low C/O ratio in those hot 
exoplanet atmospheres, even though the exact ratio cannot be determined.

Due to the presence of clouds, however, the detection of water vapor, and any 
first constraint on C/O ratios has been a very challenging problem \citep{sing:2016}. 
In fact, to date the only transiting exoplanet for which the C/O ratio has been constrained 
with this technique is WASP-12b, which has been shown to have C/O $<1$ at the 
terminator region \citep[at 3-sigma confidence][]{kreidberg:2015}. Other techniques 
have provided constrains as well. For example, the C/O ratios of the four directly 
imaged planets orbiting HR8799 have been recently estimated by \cite{lavie:2016}, 
who suggest the two inner planets (at $\sim 14$ and $\sim 25$ AU) show at least 30 times 
lower-than-stellar C/O ratios, while the two outer planets (at $\sim 38$ and $\sim 68$ AU) 
show $\sim 1.5$ larger-than-stellar C/O ratios.

\cite{oberg:2011} provided a partial framework for understanding the C/O ratio of gas 
giants within a simple disk, where the C/O ratio of the solids and gas were altered 
via condensation, which varies with the location of ice lines.  Typically, the 
condensation of H$_2$O and CO$_2$ increase the C/O ratio of the remaining gas compared 
to solar abundances, while depressing the C/O of solids. This has been taken by much 
of the community as a ``prediction" that planets that form in disks via core accretion, 
since they accrete potentially hundreds of Earth masses of gas, should be enhanced in 
C/O compared to their stars. However, this logic ignores the important, and 
potentially dominant reservoir of C and O of the solid material in the disk.

Knowledge of the total amount of solid material accreted by gas 
giant planets could lead to a better understanding of the C/O ratio of their envelopes 
and visible atmospheres. For example, if the amount of solids accreted is negligible, 
then the composition of the planetary envelope will follow the composition of the gas 
in the protoplanetary disk which can, if accreted in certain parts of the disk, allow 
for larger-than-stellar C/O ratios \citep{oberg:2011,oberg:2016}. Similarly, if we knew 
that a planetary envelope and atmosphere is polluted with a large amount of solid 
material, then the atmosphere would have a lower-than-stellar C/O ratio, due to 
the oxygen-rich nature of the solid material in comparison with the gas in the disk.

There is a sample of planets for which we can make an assessment of the amount of 
accreted solids in their H/He-dominated envelopes.  \cite{thorngren:2015} has recently 
shown, using a sample of 47 warm transiting giant exoplanets ($T_{\rm eq}<1000$K), 
whose radii are not affected by radius inflation processes, that these giant planets, 
as a class, are metal enriched in comparison to their parent stars. In particular, 
their calculations show that these planets have metal masses that range from tens 
to hundred of Earth-masses, and which therefore make an important part of the bulk 
mass of the planets.  Given that it is unlikely that all that mass is in the core, 
this suggest that the envelopes of these planets are highly enriched 
with metals, which is supported by models of planet accretion 
\citep{fortney:2013,venturini:2015,venturini:2016,mordasini:2016}. Given that solid material 
is the main deliverer of metals in planetary envelopes, the work of \cite{thorngren:2015} provides 
us then with estimates for the amount of solid material in the envelopes of these exoplanets, which 
in turn can be used to estimate the C/O ratios in their envelopes.

In this letter, we provide estimates for the C/O ratios for the sample of planets 
in \cite{thorngren:2015}. We do this by using the simple static accretion model 
of \cite{oberg:2011}, which assumes the envelope gas and solid material are accreted 
from the same region in the disk, although our main conclusions do not rest on choice of 
accretion model. We detail the model and our calculations in Section \ref{sec:estimation}, and discuss the implications for present and 
future observational studies of giant exoplanet atmospheres in Section \ref{sec:discussion}.

\section{Estimation of C/O ratios}\label{sec:estimation}

To estimate the C/O ratios of the \cite{thorngren:2015} planetary sample, 
we use the straightforward static planetary accretion model described 
by \cite{oberg:2011}.  Their framework gives the deviation of element $X$ 
from the stellar value in an accreted planetary atmosphere as 
\begin{eqnarray*}
a_X = \frac{f_{X,s}}{f_{s/g}}\frac{M_s}{M_g} + (1-f_{X,s}),
\end{eqnarray*}
where $f_{X,s}$ is the fraction of element $X$ bound in solids, $f_{s/g}$ is the 
grain-to-gas ratio in the disk (which, as in \cite{oberg:2011}, we take to be equal 
to 0.01, the observed grain-to-gas ratio in the interstellar medium), $M_s$ is the 
amount of mass in the envelope accreted in solids and $M_g$ is the amount of mass 
in the envelope accreted in gas.  With this, the deviation of the C/O ratio of the 
planet compared to the star is given by $a_C/a_O$. 

In order to compute $a_C$ and $a_O$, we follow \cite{oberg:2011} and use the data 
in their Table 1, which compiles abundances and evaporation temperatures for CO, 
CO$_2$, H$_2$O, carbon grains and silicates for conditions typical in a protoplanetary 
disk, which in turn allows us to compute the fraction of C and O bound in solids 
(i.e., $f_{X,s}$ for $X=\textnormal{C}$ and $X=\textnormal{O}$) for a given temperature, 
which we summarize in Table \ref{table:fractions}. In order to obtain the temperature at different positions 
in the protoplanetary disk, \cite{oberg:2011} use a simple disk temperature profile, 
which is a simple power-law profile of the form
\begin{eqnarray*}
T(r) = T_0 \left(\frac{r}{1\ \textnormal{AU}}\right)^{-q},
\end{eqnarray*}
where they set $T_0 = 200$ K and $q = 0.62$, which are average values for a large sample 
of protoplanetary disks \citep{AW:2007}. We use the same parametric profile here.

\begin{deluxetable}{cccc}
\tabletypesize{\footnotesize}
\tablewidth{0.1\textwidth}
\tablecolumns{4}
\tablecaption{ Fraction of solids ($f_{X,s}$) by element as a function of temperature in the protoplanetary disk \label{table:fractions}}
\tablehead{
\colhead{Element} &
\colhead{$T>135$ K} &
\colhead{$T>47$ K } &
\colhead{$T>20$ K }
}
\startdata
C & 0.25 & 0.25 & 0.375\\
O & 0.32 & 0.52 & 0.66\\
\enddata
\tablecomments{Fractions obtained using the data in Table 1 of \cite{oberg:2011}.}
\end{deluxetable}

The key ingredient needed to estimate $a_C/a_O$ for the sample of planets are bulk gas 
mass, $M_g$, and mass accreted in solids, $M_s$. In order to obtain these values 
we use the joint posterior distribution function (PDF) of the total estimated mass for each 
planet, $M_\textnormal{tot}$, and the estimated mass in metals, $M_Z$. We obtain the 
latter with the method described in \cite{thorngren:2015}. In short, the method uses 
the observed planetary masses, planetary radii, and system ages, along with giant 
planet thermal evolution / contraction models, to find solutions for the total metal 
mass with a planet that matches the measured radius at the given system age. These 
models place up to the first $10M_\Earth$ of metals in the core, while the rest of the 
metals that have to be added in order to match the observed radii are mixed in 
the (hydrogen-dominated) envelopes. In order to explore how the assumption of a 
$10M_\Earth$ core impacts our estimations, we repeat these calculations by assuming 
cores composed of $5M_\Earth$ and $15M_\Earth$, which clearly would yield either more 
or less metals in the envelope.

In order to estimate $M_s$, we note that $M_Z$ can be written as a 
contribution between the core mass, $M_c$, (whose hydrogen fraction we consider 
negligible), and the metals donated by the gas and the solids, i.e.,
\begin{equation}
\label{mz}
M_Z = M_c + Z_gM_g + Z_sM_s,
\end{equation}
where $Z_g$ and $Z_s$ are the gas and solid metal fractions, respectively. $M_\textnormal{tot}$, 
on the other hand, can be written as the contribution between $M_c$, $M_g$ and $M_s$, i.e.,
\begin{equation}
\label{mtot}
M_\textnormal{tot} = M_c + M_g + M_s.
\end{equation}
Finally, solving for $M_s$ using equations (\ref{mz}) and (\ref{mtot}), we obtain
\begin{equation}
\label{ms}
M_s = \frac{M_z-M_c}{Z_s-Z_g} - Z_g\frac{M_\textnormal{tot}-M_c}{Z_s-Z_g}.
\end{equation}  
In our calculations, we set $Z_g = 0.01$ and $Z_s = 1$ in this expression, 
which are typical values for protoplanetary disks where the gas is H/He dominated. 
Note that $M_g$ is easily obtained from these equations.

\begin{figure*}
\epsscale{1.1}
\plottwo{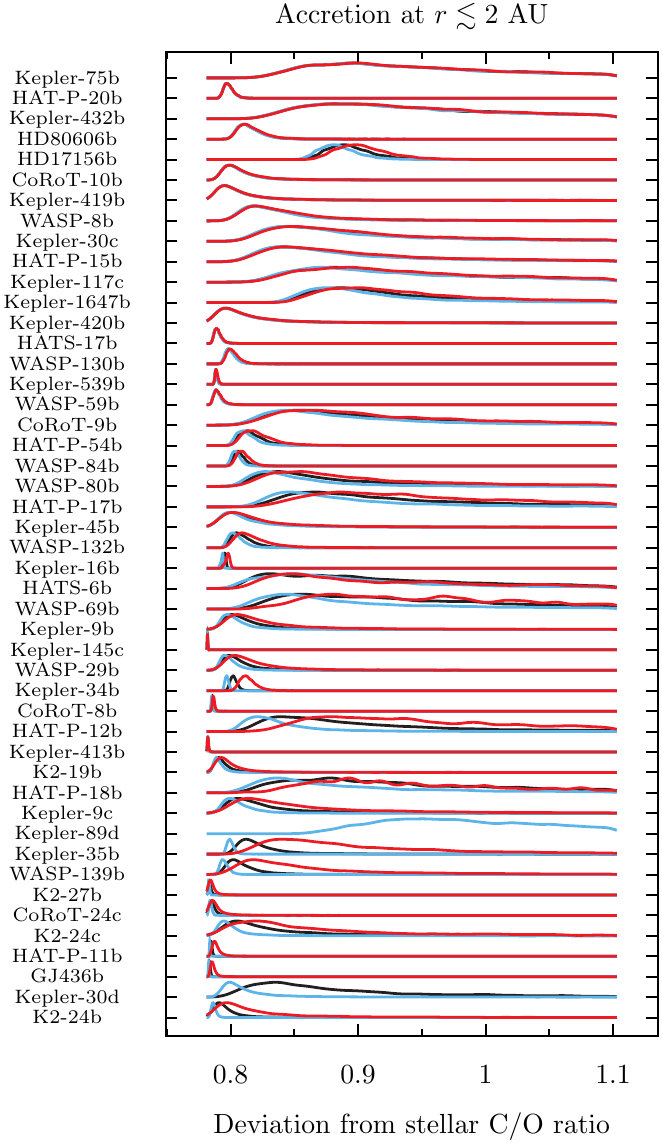}{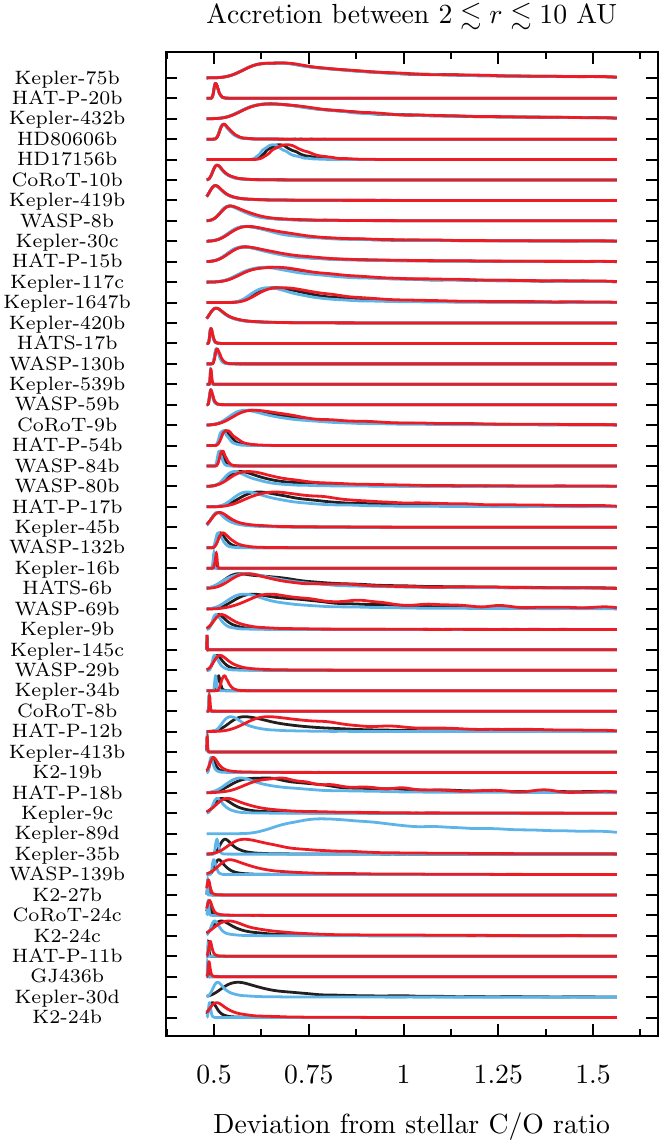}
\caption{Posterior distributions (smoothed with a Gaussian KDE for visualization) of the C/O ratio 
deviation for accretion at distances less than 2 AU and between 2 and 10 AU in the protoplanetary disk for the sample
planets of \cite{thorngren:2015}. The colors indicate different amounts of metals sequestrated in the core: $5M_\Earth$ (blue),
$10M_\Earth$ (black) and $15M_\Earth$ (red). Planets are ordered by mass, with more massive planets on top.}
\label{fig:copred1}
\end{figure*}

\begin{figure}
\epsscale{1.1}
\plotone{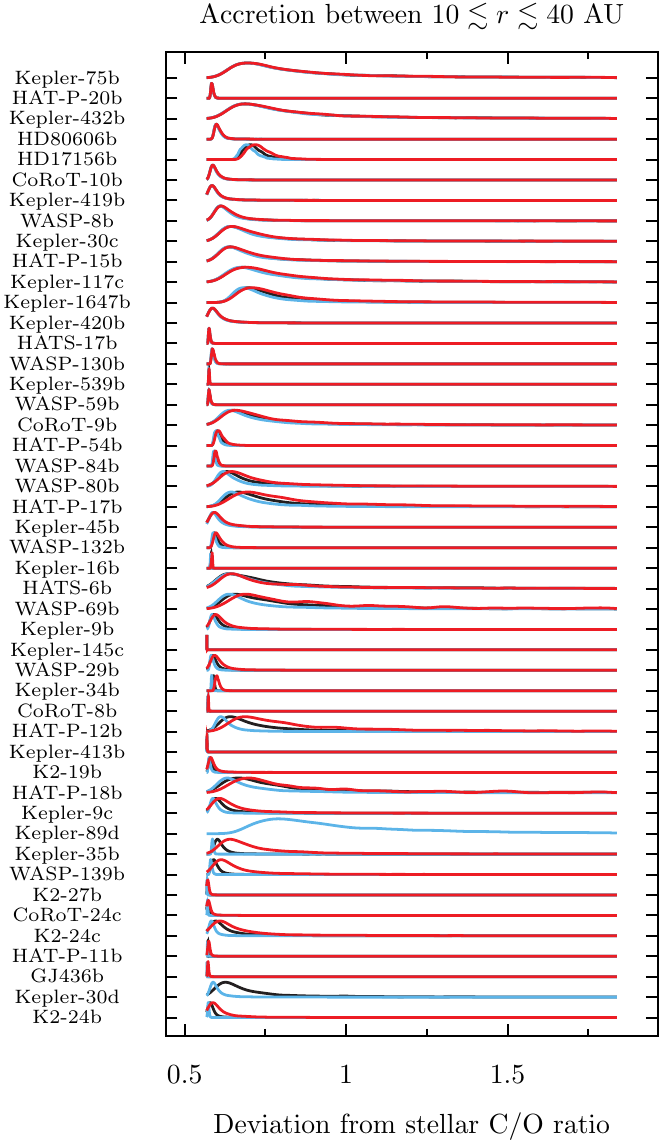}
\caption{Same as Figure \ref{fig:copred1}, but for accretion between 10 and 40 AU.}
\label{fig:copred2}
\end{figure}

10,000 draws from the joint PDF $(M_\textnormal{tot},M_Z)$ were 
obtained for each planet in order to obtain the joint PDF of 
$(M_s,M_g)$, which was used to estimate the PDF of $a_C/a_O$. 
Figure \ref{fig:copred1} and \ref{fig:copred2} show the results of this sampling scheme 
for accretion at different positions in the protoplanetary disk for planets with different core 
masses, which are color coded: $5M_\Earth$ (blue), $10M_\Earth$ (black) and 
$15M_\Earth$ (red). The planets are ordered in the figures in terms of their mass with more massive 
planets are on top.

In general, the predicted C/O ratio deviations for almost all 
planets fall \textit{under} the stellar value (=1). The reasons for strongly sub-stellar values 
are simple.  First, the amounts of metals (our proxy for the accreted solids in the 
planetary envelopes) estimated in \cite{thorngren:2015} are quite large, typically 
tens of Earth masses, or even hundreds of Earth masses for super-Jupiters, far in 
excess of ``solar composition". Thus, the C/O ratio in the solid material in 
the disk, which has a large mass fraction of C and O, overwhelms the C/O ratios 
of the accreted gas, which has little C or O, in setting the total C/O of the 
accreted envelope. This, in turn, suggests that the accretion of solids, rather 
than gas, is what define the ratios in the envelopes of most giant exoplanets, in 
agreement with recent planet formation modeling work \citep{mordasini:2016}. 

As can be observed in Figures \ref{fig:copred1} and \ref{fig:copred2}, the assumed core mass has in general a very 
small but consistent effect: a larger core mass implies a higher (more stellar or 
super-stellar) C/O ratio deviation. The effect is most noticeable for less massive planets, 
specially for the relatively low mass Kepler-30d and Kepler-89d,   
because the estimated metal masses ($5-9M_\Earth$) are not significantly larger than the core 
masses considered here. This implies that if we assume core masses of 
the same order or higher to the estimated metal masses, then the planets are consistent 
with having all the metals sequestered in their cores (which is the reason why the case in which 
we assume core masses of 15$M_\Earth$ is not shown in these figures for Kepler-30d, while the 
cases in which we assume 10 and 15$M_\Earth$ cores are not shown for Kepler-89d). If this indeed 
was the case, then larger-than-stellar C/O ratios should be expected on the envelopes of those 
planets.

\begin{figure}
\epsscale{1.1}
\plotone{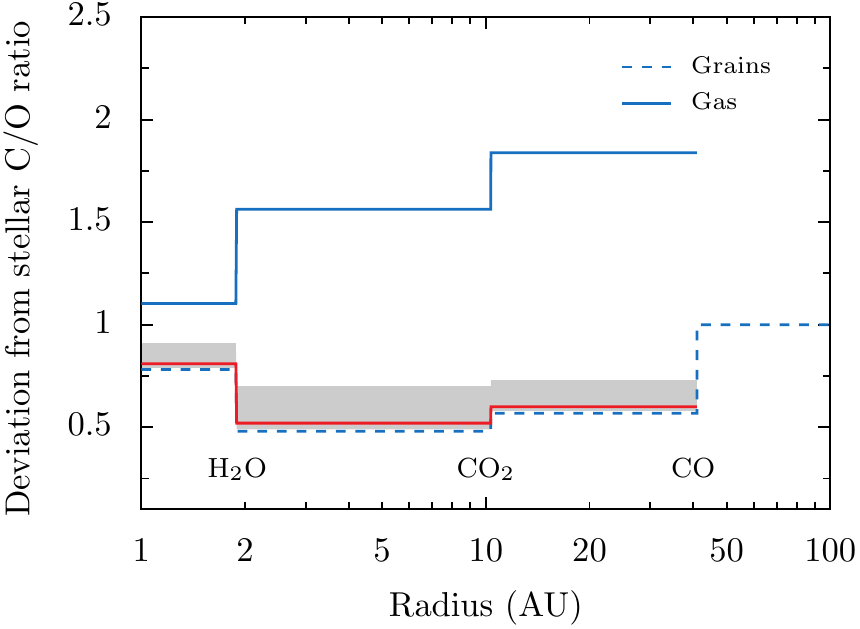}
\caption{C/O ratio deviation in the (mean) protoplanetary disk used in this work. Blue lines denote the 
deviation in the gas (solid lines) and grains (dashed lines) in the disk, while the red lines with gray bands show 
the deviation of the sample of planets studied in this work. Ice lines of the different elements considered are 
indicated in the figure.}
\label{fig:coall}
\end{figure}

It is interesting to note that the deviations from the stellar C/O ratios of almost 
all the planets peak at (or very close to) the deviation from stellar C/O ratio 
of the solid material, which is the deviation expected when $M_s/M_g >> f_{s/g}$; in this 
limit, $a_C/a_O\approx f_{C,s}/f_{O,s}$. For the 
case of accretion at distances smaller than $2$ AU (i.e., inwards to the H$_2$O ice-line), 
the C/O ratio of the solid material is $0.78$, for accretion between 
$2\lesssim r \lesssim 10$ AU (i.e., between the H$_2$O and CO$_2$ ice-lines) is 
$0.48$ and for accretion between $10\lesssim r \lesssim 40$ AU (i.e, between the 
CO$_2$ and CO ice-lines) is $0.57$; as can be seen in Figures \ref{fig:copred1} 
and \ref{fig:copred2}, most distributions of the C/O ratio deviations for the planets in 
our sample peak closer to those values. In Figure \ref{fig:coall} we plot the 
median deviation for all the planets studied in this work along with the 16th and 84th percentile 
of the sample (for the case in which we assumed a $10M_\Earth$ core) along with the C/O ratio 
deviation of the gas and solid material for the different distances studied here. This 
nicely illustrates the fact that, in our sample, almost all the planets show C/O ratio deviations 
closer to that of the solid material. 

\section{Discussion}\label{sec:discussion}

In the present work, we have estimated C/O ratio deviations for the 47 warm giant 
exoplanets studied in \cite{thorngren:2015}, and for which estimates of the amount 
of metals, here used as a proxy of the solid material in the envelopes, is available. 
Using a simple static model of planetary accretion, we estimate that most planets have 
lower than stellar C/O ratios in their envelopes, given their overall strongly 
metal-enriched nature.

If we couple this result with the observational determination that most stars 
have C/O ratios lower than 1 \citep{brewer:2016,teske:2014,nissen:2014}, 
then this implies that we estimate C/O $<1$ for most of the planets in the sample 
of \cite{thorngren:2015}, independent of where the planet formed as long as (1) 
it was formed inwards to the CO ice-line \citep[which is the expected formation location 
in the core accretion scenario, see, e.g., ][and references therein]{lissauer93,armitage2007}, 
(2) the envelope is well mixed with minimal metallicity gradient \citep[expected for the masses of the planets being considered 
in this work, see ][]{mordasini:2016} and (3) the accreted solids are oxygen-rich in relation 
with the gas \citep[which is expected in typical conditions in disks, see, e.g., ][]{oberg:2011, 
pontoppidan:2006}. 

The atmospheric characterization of warm exoplanets like the ones studied in this work 
via spectroscopy would be ideal in order to test our findings. However, due to their cool atmospheres 
and corresponding small scale-heights, and their potentially cloudy nature, the 
constrains that can be put on their C/O ratios today is quite challenging \citep[see, e.g., the case of 
HAT-P-11b][]{fraine:2014}. If hotter giant exoplanets follow similar formation 
paths as their warm counterparts, however, this means that most of them will also 
have C/O ratios of less than one, which would in turn imply that water features 
should be ubiquitous in their atmospheres if equilibrium conditions 
hold \citep{madhusudhan:2012,kreidberg:2015}, as long as their atmospheres are 
clear of clouds and are representative of the bulk C/O ratios in their envelopes. 
Low C/O ratios have indeed been constrained in HD 209458b
\citep{brogi:2016} and WASP-12b (under equilibrium conditions, \citealp{kreidberg:2015}; but see \citealp{stevenson:2014,hansen:2014}); 
however, for other exoplanets it is unclear what constraints do the observed water signatures impose on their water abundances \citep{sing:2016, barstow:2017}. 
Brown dwarfs spectra could also provide a useful comparison sample to our giant planet population; however, it is unclear if they form in disks 
around stars. The very wide diversity in C/O ratios obtained for T dwarfs \citep{line:2016,madhu:2016} 
could be due to chemical sequestration of O in condensates, a variety of formation mechanisms for the 
objects, or imperfections in opacity databases.

\begin{figure}
\epsscale{1.1}
\plotone{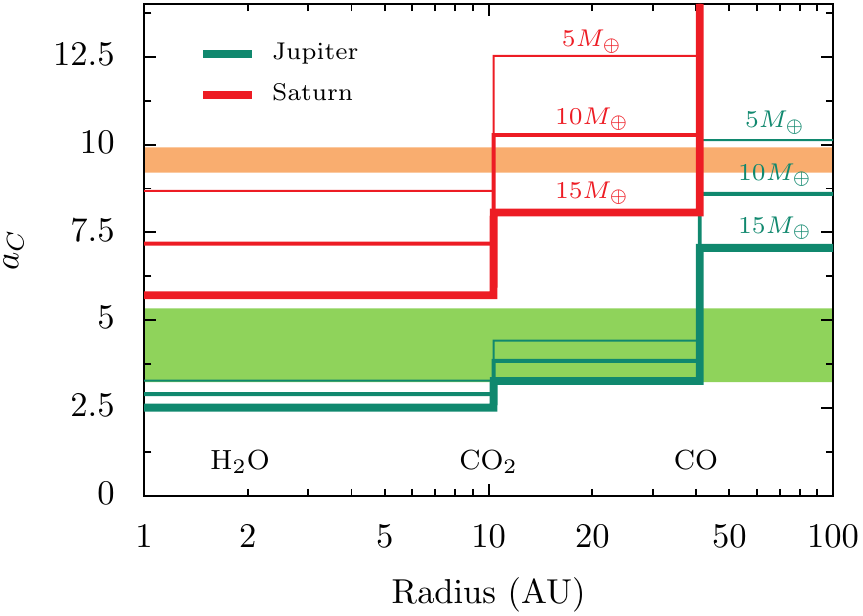}
\caption{Comparison between estimates from our model for $a_C$ (solid lines, line tickness indicates different assumed core masses) 
and empirical estimates from \citet[][Saturn, orange band; Jupiter, green band]{atreya:2016}. Ice lines of the different elements 
considered are also indicated.}
\label{fig:ac_planets}
\end{figure}

It is interesting to note that our model works nicely for our own Solar System giant planets, Jupiter and Saturn. In 
Figure \ref{fig:ac_planets} we show estimations for $a_C$ (C/H) using our model for these planets assuming different core masses 
to obtain $M_Z$, as in Section \ref{sec:estimation}, and compare them to the values estimated in \cite{atreya:2016} using data from the Galileo and 
Cassini missions. As can be seen, there is good agreement within the error bars and the uncertainty given by the core mass in our 
models. We can also use our model to estimate Jupiter's O/H, which is timely as NASA's Juno mission is going to measure this value 
soon \citep{janssen:2005, helled:2014}. Our model suggests $a_O$ values of 2.9-3.9, 4.1-5.7 and 5.0-7.0 for formation inside the H$_2$O, 
between the H$_2$O and CO$_2$ and between the CO$_2$ and CO ice-lines respectively; slightly larger than the ones predicted in, e.g., 
\cite{mousis:2012} and \cite{wang:2015}. These values should be taken with care, as (1) Jupiter's internal structure (and, thus, $M_Z$) depends on 
the equations of state and the accuracy of the observational constrains (e.g. 
gravitational moments) \citep[see, e.g.,][and references therein]{saumon:2004,miguel:2016} and (2) the distribution of $M_Z$ in Jupiter might be 
inhomogeneous.

Several additional steps could be taken if one wanted to perform a more 
detailed estimation of the C/O ratios of the studied planets. First, the abundances of carbon and oxygen bearing elements, 
which were obtained from the work of \cite{pontoppidan:2006} might change from disk to disk. Second, radial drift 
of the material in the disk and gas accretion should be taken into account in order to 
estimate how the snowlines and the abundances of the different elements that define the 
changes in the C/O ratios in the disk at the distances proved here (i.e., H$_2$O, CO$_2$, CO) evolve in the 
disk \citep{piso:2015}. Third, the (planetary and disk) time evolution with respect to variables 
such as pressure and temperature should be considered \citep{mordasini:2016}. The effect of implementing 
all these more detailed processes would be, in general, to shift the ice-lines inwards in the protoplanetary disk. 
Furthermore, although the relative ammount of oxygen and carbon in solids and gas might change, the fact that solids 
are oxygen rich seems to be a reasonable assumption \citep{mordasini:2016,wilson:2016}. The detailed impact 
of these effects is outside the scope of this work, whose aim was to show an alternative interpretation to the 
typical picture of envelope accretion in protoplanetary disks, in which gas accretion 
is suggested to be the main character that define important abundance tracers such as the 
C/O ratio in planetary envelopes \citep[e.g., ][]{oberg:2011,cridland:2016,oberg:2016,madhusudhan:2016}. 
Given that giant planets are heavily metal enriched compared to their parent stars, we suggest that it is in 
fact the solids, and not the gas, that are the key ingredient that define these tracers in the 
envelopes of giant exoplanets. 

\acknowledgments

We thank an anonnymous referee for useful suggestions that improved this article. N.E. and Y.M. acknowledge support 
from the Kavli Summer Program in Astrophysics. N.E. acknowledges 
support from VRI/PUC, CONICYT-PCHA/Doctorado Nacional and the Ministry for the Economy, Development and Tourism Programa Iniciativa Cient\'ifica
Milenio through grant IC 120009, awarded to the MAS. J. J. F. acknowledges 
the support of NSF grant NNX16AB49G. Y.M. greatly appreciates the CNES post-doctoral fellowship program. N.E. would like to thank R. Brahm and A. Jord\'an for useful 
discussions.

\allauthors

\listofchanges

\end{document}